\documentclass[prd,twocolumn,amsmath,amssymb,eqsecnum]{revtex4-1}
\pdfoutput=1
\usepackage{footnote,verbatim,slashed,graphics,graphicx,color,slashed,textcomp}

\graphicspath{{figures/}}

\usepackage[colorlinks=true,
linkcolor=red,
urlcolor=blue,
citecolor=blue]{hyperref}

\newcommand{\GeV}{\mathrm{GeV}}
\newcommand{\TeV}{\mathrm{TeV}}

\begin{document}
	
\title{Pure Gravitational Dark Matter, Its Mass and Signatures
	%Dark Matter with Effective Theory of Gravity
	}
\author{Yong Tang}
\email{ytang@kias.re.kr}
\affiliation{Korea Institute for Advanced Study, Seoul 02455, South Korea}
\author{Yue-Liang Wu}
\email{ylwu@itp.ac.cn}
\affiliation{State Key Laboratory of Theoretical Physics(SKLTP) \\
	Kavli Institute for Theoretical Physics China (KITPC) \\
	Institute of Theoretical Physics, Chinese Academy of Sciences, Beijing 100190 \\
	University of Chinese Academy of Sciences (UCAS), P. R. China}

\begin{abstract}
In this study, we investigate a scenario that dark matter (DM) has only gravitational interaction. In the framework of effective field theory of gravity, we find that DM is still stable at tree level even if there is no symmetry to protect its longevity, but could decay into standard model particles due to gravitational loop corrections. The radiative corrections can lead to both higher- and lower-dimensional effective operators. We also first explore how DM can be produced in the early universe. Through gravitational interaction at high temperature, DM is then found to have mass around $\TeV \lesssim m_X\lesssim 10^{11}\GeV$ to get the right relic abundance. When DM decays, it mostly decays into gravitons, which could be tested by current and future CMB experiments. We also estimate the resulting fluxes for cosmic rays, gamma-ray and neutrino.
\end{abstract}

\maketitle 

\section{Introduction}
Evidence for the existence of dark matter (DM) is compelling, supported from astrophysical length to cosmological scale. Despite its convincing inferences in cosmic microwave background (CMB), big-bang nucleosynthesis, large scale structure (LSS) and other astrophysical observables, DM's particle identity is still a mystery since all the confirmed evidence simply suggests DM should have gravitational interaction. 

From the current experimental searches for DM, we have already known that the interaction between DM and the standard model particle should be weak. DM is stable or at least has a very long lifetime, much longer than the age of our Universe $t_U\sim 10^{17}$s, otherwise it can give rise observable signatures in CMB, LSS, cosmic rays, gamma-ray and neutrino experiments. 

It is a logical possibility that DM might have only gravitational interaction. It is usually expected there would be no way to produce DM since its interaction with standard model (SM) particle is super weak, not to mention how to search for it. This is true if we only consider classical theory of gravity where DM is stable even though there might be no symmetry to protect its longevity. However, as we shall show in this paper, if perturbative gravitational loop corrections are taken into account, effective operators are induced and can make DM decay.

To estimate the quantum correction from graviton, a consistent quantum theory of gravity is needed. For our purpose in this study, we may investigate properties of DM within the framework of effective field theory of gravity~\cite{Burgess:2003jk,Donoghue:1994dn}.  This framework is justified if we only consider low-energy processes or weak gravity approximation which are satisfied in the problem of our interest. Recently, it has been shown that the general theory of relativity can be derived as an effective field theory of gravitational quantum field theory with spin and scaling gauge symmetries~\cite{wu:2015wwa}.

This paper is organized as follows. In Sec.~\ref{sec:GDM} we establish the conventions and definitions. Then in Sec.~\ref{sec:prod}, we explore a possible production mechanism of DM and obtain its viable mass range. In Sec.~\ref{sec:EffoP}, we illustrate how effective operators that induce interactions between DM and SM particles could arise. In Sec.\ref{sec:SGDM} we present possible decay channels for DM and its signatures. Finally, we give a summary.

\section{Gravitational Dark Matter (GDM)}\label{sec:GDM}
We start with a minimal setup with one scalar dark matter (DM) field $ X$ and one non-DM scalar $\phi$.  Here $\phi$ may be  denoted collectively as any non-DM fields. Explicit extensions with fermionic and gauge fields will be discussed later. In the flat spacetime, the general action $\mathcal{S}$ would be an integration over Lagrangian density $\mathcal{L}$,
\begin{align}\label{eq:lag}
\mathcal{L}=\frac{1}{2}\partial _\mu \phi \partial^\mu \phi + \frac{1}{2}\partial_\mu X \partial^\mu X -\mathcal{V}\left(\phi, X\right),
\end{align}
We explore our investigation with the following potential,
\begin{align}
\mathcal{V}\left(\phi,  X\right) & = V_\phi\left(\phi \right)+V_X\left(X \right),\\
V_\phi\left(\phi\right) &= \frac{1}{2}m^2_\phi\phi^2 
%+ \frac{1}{3!}m_\phi \mu_{\phi} \phi^3 
+ \frac{1}{4!}\lambda_{\phi} \phi^4, \\
V_X   \left(X   \right) &\supset \frac{1}{2}m^2_X  X^2 .
%+ \frac{1}{4!}\lambda_{X} X^4. 
\end{align}
where $m_{i}$ and $\lambda_i$ are masses and quartic couplings, respectively. 
%In the potential, we have not included linear term $\sim \phi$ which can be translated away. 
There is a discrete $Z^X_2$ symmetry, $ X \rightarrow -  X$, which can protect the stability of DM $X$. When there is no any direct or indirect interactions other than gravity between $X$ and $\phi$, we shall refer $X$ as {\textbf{Gravitational Dark Matter}} (GDM). 

In flat Minkowski spacetime without gravity, $\phi$-$ X$ system with potential $\mathcal{V}$ are renormalizable in the sense that all ultraviolet (UV) divergences from loop corrections can be absorbed into fields, masses and couplings, and that no other counterterm needs to be introduced. So $\phi$ and $X$ are completely decoupled at all scale even if after renormalization group flow. However, as we shall see soon, the above picture will be changed dramatically after including perturbative gravitational effects and non-zero terms, such as $\lambda_{ X \phi} \phi^2 X^2$, can be induced, with Planck scale suppressed $\lambda_{ X\phi} \sim m^2_{ X}m^2_{\phi}/M^4_{P}$.

Now let us include gravity with standard Hilbert-Einstein action. The Lagrangian would be modified to
\begin{align}\label{eq:laggravity}
	\mathcal{L}=\sqrt{-g(x)}\Big{[}&\frac{1}{16\pi G}R+\frac{1}{2}g^{\mu \nu }\partial _\mu \phi \partial_\nu \phi \nonumber \\
	& 	+ \frac{1}{2}g^{\mu \nu }\partial_\mu  X \partial_\nu  X -\mathcal{V}\left(\phi,  X\right)\Big{]},
\end{align}
where $R$ is the Ricci scalar, $G$ is the Newton's constant, $g(x)$ is the determinant of spacetime metric tensor $g_{\mu\nu}$ and $g^{\mu\nu}$ is the inverse matrix of $g_{\mu\nu}$ with $g^{\mu\rho}g_{\rho\nu}=\delta^{\mu}_{\nu}$,
 \[
 \delta^{\mu}_\nu= \bigg{\{}
 \begin{array}{lr}
 1,& \mu =\nu \\
 0.& \mu \neq \nu
 \end{array}
 \]
Note that our framework is similar to Ref.\cite{Garny:2015sjg}, but different from those~\cite{Ren:2014mta, Cata:2016dsg} where non-minimal coupling between $R$ and $X$ is introduced, and also different from Refs.~\cite{Chung:1998ua, Kuzmin:1998kk} which rely on dynamics of quantum field theory on curved background spacetime$^{\textrm{\cite{Khlopov}}}$.

We consider the weak gravity case and express the metric field around the flat Minkowski background spacetime as follows, 
\begin{equation*}
g_{\mu\nu}=\eta_{\mu\nu}+\kappa\ h_{\mu\nu},\;\eta_{\mu\nu}=\eta^{\mu\nu} \equiv (1,-1,-1,-1),
\end{equation*}
where $ \kappa = \sqrt{16\pi G}\equiv 1/M_P$ and $h_{\mu\nu}$ is identified as quantum field for spin-2  massless graviton, propagating in flat background spacetime. This expression is useful and justified when we are only interested in environment without strong gravity and in low-energy physics if the energy is smaller than Planck scale~\cite{Burgess:2003jk,Donoghue:1994dn}.

The above expression of $g_{\mu\nu}$ is general, but the expansions of inverse metric and determinant are approximate with ignoring higher-order terms. For our purpose, it is enough to keep terms up to second order in $\kappa$ only,
\begin{align*}
g^{\mu\nu}   &= \eta^{\mu\nu}-\kappa h^{\mu\nu}+\kappa ^2
h^{\mu}_{\alpha}h^{\alpha\nu}+...,\\
\sqrt{-g(x)} &= 1+\frac{1}{2}\kappa h-\frac{1}{4}\kappa ^2 (h^{\mu\nu}h_{\mu\nu}-\frac{1}{2}h^2)+... ,
\end{align*}
where $h\equiv \eta^{\mu\nu} h_{\mu\nu}$. At this stage, we have already seen that there are infinite operators in the expansion series, which partially shows the non-renormalizability of gravity. This is not a problem in effective field theory where one can only keep terms up to $\kappa^n$ and $n$ is determined by the concerned precision. To quantize $h_{\mu\nu}$, we need to fix the gauge. We choose the harmonic gauge-fixing condition,
\begin{equation}
C^{\mu}=\partial_{\nu}h^{\mu\nu}-\frac{1}{2}\partial^{\mu}h^{\nu}_{\nu}=0,
\end{equation}
then we have the graviton propagator with a simple form in momentum space,
\begin{equation}
G^{\mu\nu\rho\sigma}(k)= \frac{i}{k^2}\left[\eta^{\mu\rho}\eta^{\nu\sigma}+\eta^{\nu\rho}\eta^{\mu\sigma} -\eta^{\mu\nu}\eta^{\rho\sigma}\right].
\end{equation}
Note that the corresponding ghost in this gauge is irrelevant for our calculations of one-loop gravitational corrections, which is similar to quantum electrodynamics with Feynman gauge. The Lagrangian now can be rewritten as
\begin{align}\label{eq:lag1}
\mathcal{L}=&\frac{1}{2}\partial _\mu \phi \partial^\mu \phi  + \frac{1}{2}  \partial_\mu  X \partial^\mu  X -\mathcal{V}\left(\phi,  X\right) \nonumber \\
&+\frac{1}{2}h\partial^2h-\frac{1}{4}h^{\mu\nu}\partial^2h_{\mu\nu}
  + \delta\mathcal{L}\left(h_{\mu\nu},  X,\phi \right),
\end{align}
where $\delta\mathcal{L}\left(h_{\mu\nu},  X,\phi \right)$ at order of $\kappa^2 $ is given by
\begin{align}\label{eq:deltalag}
& \kappa\left[\left(\frac{1}{2}h\eta^{\mu\nu}-h^{\mu\nu}\right)\left(\frac{1}{2}\partial _\mu \phi \partial_\nu \phi + \frac{1}{2}\partial _\mu  X \partial_\nu X\right) - 
\frac{1}{2}h\mathcal{V}\right] \nonumber \\
&- \frac{\kappa^2}{4} \left(h^{\alpha\beta}h_{\alpha\beta}-\frac{1}{2}h^2\right)\mathcal{V}
 + \kappa^2 \left(\frac{1}{2}\partial _\mu \phi \partial_\nu \phi + \frac{1}{2}\partial _\mu  X \partial_\nu X\right) \nonumber  \\
&\left[h^{\mu}_{\alpha}h^{\alpha\nu}-\frac{1}{2}hh^{\mu\nu}-\frac{\eta^{\mu\nu}}{4} \left(h^{\alpha\beta}h_{\alpha\beta}-\frac{1}{2}h^2\right)\right].
\end{align}
The above formulas would be the main Lagrangian or framework for discussions in next section. At this point, it is easy to check that $X$ is still stable at tree level. Even if we include a $Z^X_2$ symmetry breaking term in the potental $V_X$, 
\begin{equation}\label{eq:cubic}
V_X \supset \frac{1}{3!}m_X\mu_XX^3,
\end{equation}
$X$ is still stable since there is no available interactions to decay through. However, as we shall show in Sec.~\ref{sec:EffoP}, once we take loop corrections into account, $X$ shall decay.

The above cubic term can be induced if the discrete $Z^X_2$ symmetry is spontaneously broken. For example, if a scalar $\mathcal{X}$ has a potential
\begin{equation}
V_\mathcal{X}=-\frac{1}{2}\mu^2\mathcal{X}^2+\frac{1}{4!}\lambda \mathcal{X}^4,
\end{equation}
with $\mu>0$. Then $\mathcal{X}$ would get a vacuum expectation value, $\langle \mathcal{X}\rangle=\sqrt{6\mu^2/\lambda}\equiv v_\mathcal{X}$. Substitute $\mathcal{X}=v_\mathcal{X}+X$, we get the potential for $X$,
\begin{align}
V_X&\supset\frac{1}{6}\lambda v^2_\mathcal{X}X^2 + \frac{1}{6}\lambda v_\mathcal{X}X^3 + \frac{1}{4!}\lambda X^4\nonumber \\
&=\frac{1}{2}m^2_XX^2 + \frac{1}{3!} \mu_X m_X X^3 +\frac{1}{4!}\lambda X^4,
\end{align}
with $m_X=\sqrt{\lambda v^2_\mathcal{X}/3}$ and $\mu_X=\sqrt{3\lambda}$.
%Throughout our later discussions, we do not specify the origin of this symmetry breaking term, but only focus on its phenomenological consequences. 

\section{Production Mechanism for GDM}\label{sec:prod}
In this section, we discuss how to produce dark matter particle in the early Universe. The dominant contribution to produce gravitational dark matter is through the tree-level s-channel process by mediating a graviton, $\phi + \phi\rightarrow X + X$ ($\phi$ can be any other particle in the thermal bath). The production cross section from interactions in Eq.~\ref{eq:deltalag} has the following form
\begin{equation}\label{eq:thermalsv}
\langle \sigma v \rangle \sim \kappa^4 T^2,
\end{equation}
where $T$ is the temperature of thermal bath in the universe. Due to the weakness of gravity, the interacting rate $n_\phi \langle \sigma v \rangle$ ($n_\phi\simeq T^3$ is number density of $\phi$) is much smaller than the expansion rate of universe, so $X$ is not in thermal equilibrium with other particle.

Now let us calculate how much $X$ can be produced. The Boltzmann equation~\cite{book} that describes the changes of number density $n_X$ is given by
\begin{equation}
\frac{d\left(n_X a^3\right)}{a^3dt} =  \frac{dn_X}{dt}+3\mathcal{H}n_X=\langle \sigma v \rangle \left[n^2_X-\left(n_{\textrm{eq}}\right)^2\right],
\end{equation}
where $a$ is the scale factor, $\mathcal{H}\equiv \dot{a}/a$ is the Hubble parameter, $\sim T^2/M_P$, $n_{\textrm{eq}}$ is the equilibrium number $n_{\textrm{eq}}\sim T^3$. Define the yield $Y\equiv n_X/s$, $s$ is the entropy density, we have 
\begin{equation}\label{eq:yield}
\frac{dY}{dT}=\frac{-\langle \sigma v \rangle s}{\mathcal{H} T}\left(Y^2-Y^2_{\textrm{eq}}\right)\simeq \frac{n_{\textrm{eq}} \langle \sigma v \rangle }{\mathcal{H} T}Y_{\textrm{eq}},
\end{equation}
where we can ignore $Y$ in the right-handed side of the first equation due to $ Y\ll Y_{\textrm{eq}}$. Therefore $Y$ would be a power-law function of $T$ with positive index after we put in $\langle \sigma v \rangle \sim \kappa^4T^2$ and $n_{\textrm{eq}}\sim T^3$.  We also should sum over all particles with gravitational interaction, which means we can replace $Y_{\textrm{eq}}$ with $\sim 1$. Integrate Eq.~\ref{eq:yield} over $T$ from $\mathcal{O}\left(m_X\right)$ to the maximal temperature $T_{\textrm{max}}$, then to get the right relic abundance of $X$, we would need
\begin{equation}\label{eq:abundance}
\left.\frac{n_\phi \langle \sigma v \rangle }{\mathcal{H}}\right|_{T=T_{\textrm{max}}}\simeq Y_{X}\equiv\frac{\Omega_X m_p}{\Omega_b m_X}\eta,
\end{equation}
where $T_{\textrm{max}}$ may refer as the maximal temperature of our universe after inflation, or reheating temperature, $\Omega_b$ and $\Omega_X$ are the energy density fractions of baryon and dark matter, respectively, $\Omega_X/\Omega_b\simeq 5$, $m_p\simeq 1\GeV$ is proton mass and $\eta\simeq 6\times 10^{-10}$ is baryon-to-photon ratio. From Eq.~\ref{eq:abundance}, we obtain
\begin{equation}\label{eq:relic}
m_X \sim \frac{\Omega_X M^3_P}{\Omega_b T^3_\textrm{max}}m_p\eta. 
\end{equation}
We also need check whether the universe can be hot enough to produce $X$, namely $T_{\textrm{max}}\gtrsim \mathcal{O}\left(m_X\right)$. With Eq.~\ref{eq:relic} we have
\begin{equation}
T_{\textrm{max}} \gtrsim  \left(\frac{\Omega_X M^3_P m_p\eta }{\Omega_b}\right)^{1/4}\simeq
10^{-7}M_P, 
\end{equation}
which gives the lower bound for $T_{\textrm{max}}$ and can be definitely satisfied without violating any experimental limits.
The reason why we take $T_{\textrm{max}}\gtrsim \mathcal{O}\left(m_X\right)$ as a constraint is that when the temperature is much lower than $m_X$ there would be an exponential suppression of particles that are energetic enough to produce DM $X$. Only particles at the very high energy tail of Bose-Einstein/Fermi-Dirac distributions can contribute to the production. 

The upper bound for $ T_{\textrm{max}}$, on the other hand, depends on the details of cosmic evolution. It is reasonable to expect that DM $X$ is mostly produced after inflation since otherwise it would be extremely diluted by the exponential expansion. If we take $ T_{\textrm{max}}$ as high as the inflation scale $\lesssim 10^{-4}M_P$ which is constrained by non-observation of primordial gravitational wave~\cite{Array:2015xqh}, we can have an upper bound on $T_{\textrm{max}}\lesssim 10^{-4}M_P$.

With the above bound $10^{-7}\lesssim T_{\textrm{max}}/M_P\lesssim 10^{-4}$, then from Eq.~\ref{eq:relic}, we would have a finite range for the mass of gravitational dark matter,
\begin{equation}\label{eq:massrange}
10^3m_p\lesssim m_X\lesssim 10^{-7}M_P,\TeV \lesssim m_X\lesssim 10^{11}\GeV.
\end{equation}
This is one of our main results, which predicts the mass range for GDM.  This result is consistent with the case $m_X<T_{\textrm{max}}$ in \cite{Garny:2015sjg} which also considered $T_{\textrm{max}}>m_X$ case, in the context of reheating process. Note that the above discussions do not depend on whether GDM is a scalar~\cite{Nurmi:2015ema}, fermion or vector. In later sections, we shall focus on scalar case for further investigations.

\section{Effective Operators Out of Gravity }\label{sec:EffoP}

\begin{figure}
	\begin{center}
		\includegraphics[scale=0.5]{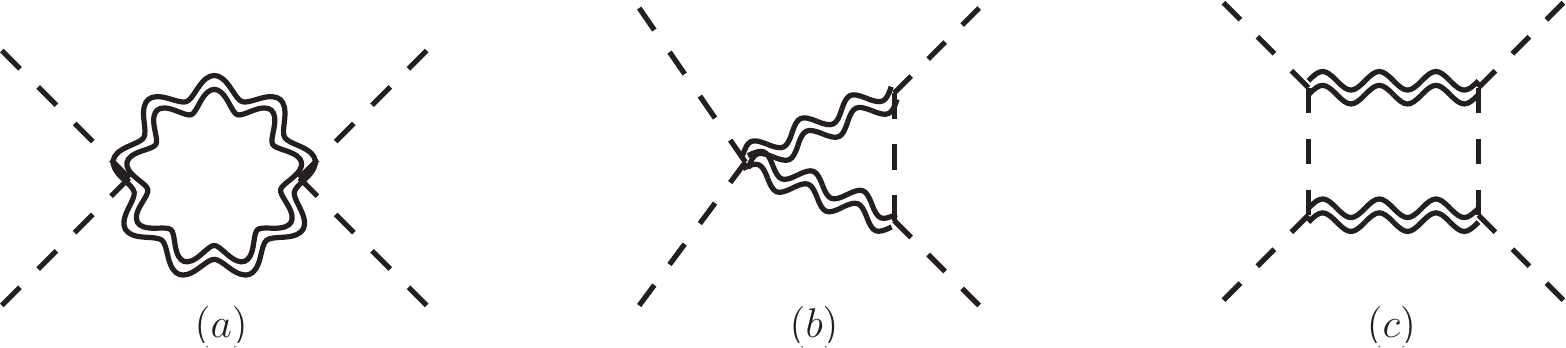}
	    \includegraphics[scale=0.5]{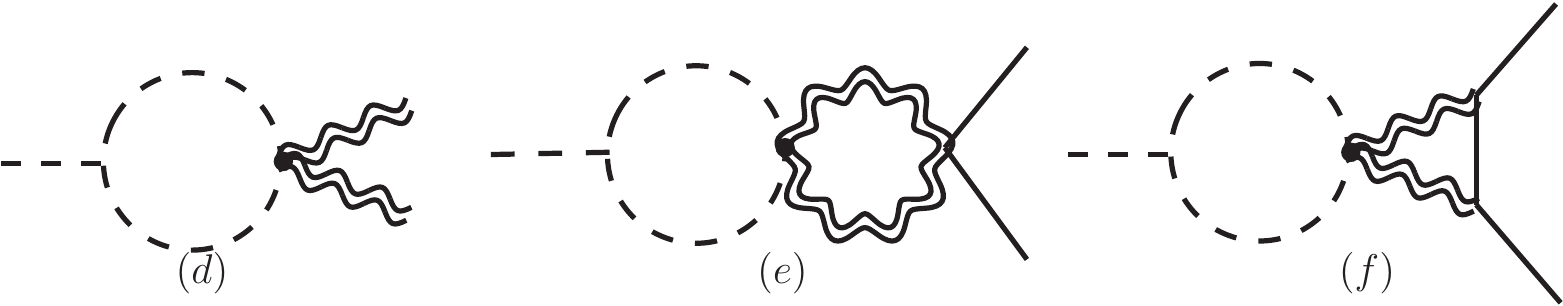}
		\caption{\label{fig:feynman}Some typical Feynman diagrams at loop level. Scalars (gravitons) are displayed with dashed (double) lines. Solid lines can be scalar, fermion and gauge fields.}
	\end{center}
\end{figure}

In this section, we consider loop contributions and show how they induce both lower- and higher-dimensional operators. Some typical Feynman diagrams are shown in Fig.~\ref{fig:feynman}. Calculations of these diagrams are involved with UV divergences which can be handled with regularization. To cancel these UV divergences, we must introduce counterterms 
and perform renormalization. This is equivalently to say that, renormalization group flow shall necessarily introduce non-zero coefficients for the operators that are allowed by the symmetry.
Below we just list some operators with explicit mass-dependent coefficients,
\begin{align}
& m^2_\phi m^2_X X^2\phi^2, m^4_X X^4, m^4_\phi\phi^4,  
m^2_X X^2 \partial^\mu \phi \partial _\mu \phi, \nonumber \\
&m^2_\phi\phi^2 \partial^\mu X \partial_\mu X,   \lambda_\phi m^2_X X^2 \phi^4, 
\cdots ,
\end{align}
with common factor $\kappa^4/16\pi^2$. The finite coefficients $\mathcal{C}_{i}$ in front of these operators are of the following form,
\begin{equation}
\mathcal{C}_{i}\sim \mathcal{O}\left(\ln\frac{\mu^2}{E^2}\right)+...,
\end{equation}
after we introduced the counterterms to cancel the divergences, where $\mu$ is the renormalization scale, $E$ is the energy, $``..."$ refers to finite $\mathcal{O}(1)$ constant. 

Note that discrete $Z^ X_2$ symmetry is still maintained if $\mu_X=0$ in the potential $\mathcal{V}$. 
%Graviton's corrections to $X^4$ and $\phi^4$ will contribute to the renormalization of $\lambda_X$ and $\lambda_\phi$. 
Importantly, as seen above, non-zero $\lambda_{ X \phi}\phi^2X^2$ term is induced with coupling
\begin{equation}
\lambda_{ X \phi }\sim \kappa^4\frac{m^2_\phi m^2 _ X}{16\pi^2},
\end{equation}
which is inevitable once we include gravity. In principle, those induced operators would also contribute to the production of GDM in the early universe. However, their contributions are further suppressed by $\kappa^4m^4_X$, in comparison with the leading one in Eq.~\ref{eq:thermalsv}. 

Now let us discuss the case without discrete $Z^ X_2$ symmetries or $\mu_X \neq 0$ in Eq.~\ref{eq:cubic}. At two-loop level, as shown in Fig.~\ref{fig:feynman} (e) and (f), we have operators as following,
\begin{align}\label{eq:decay1} 
m^2_\phi X\phi^2,  
X \partial^\mu \phi \partial _\mu \phi,
\end{align}
with a common factor, $\kappa^4\mu_X m_X^3/256\pi^4 $. Both operators are not present in the starting Lagrangian and can lead $X$ to decay into two $\phi$s if $m_X>2 m_\phi$. $X$ can also decay into two gravitons due to one-loop diagram from Fig.~\ref{fig:feynman} (d) with effective operators like 
\begin{equation}\label{eq:decay2}
	 Xh^{\alpha\beta}h_{\alpha\beta},Xh^2, 
\end{equation}
with prefactor $\kappa^2\mu_X m_X^3/16\pi^2$. 
These two operators could arise from covariant term $\sqrt{-g}X R$ after renormalization and using equation of motion for $X$.

We are now in a position to discuss the connections between GDM and SM particles. The Lagrangian can be written collectively as
\begin{align}\label{eq:smEFT}
&\mathcal{L}=\sqrt{-g}\left[\frac{R}{16\pi G} + \frac{1}{2}g^{\mu \nu }\partial_\mu  X \partial_\nu  X -\mathcal{V}\left(X\right)\right]+\mathcal{L}_{\textrm{SM}},\nonumber \\
&\frac{\mathcal{L}_{\textrm{SM}}}{\sqrt{-g}}=\left[\bar{\psi}(i\slashed{D}-m_\psi)\psi-\frac{1}{4}F_{\mu\nu}F^{\mu\nu} -\frac{1}{\sqrt{2}}(y\bar{\psi}\psi \phi + h.c.) \right.\nonumber\\
 &\left. + \frac{1}{2}D_\mu \phi D^\mu \phi -\mathcal{V}(\phi)\right], 
\end{align}
where $D$ is the covariant derivative, and in the second line, terms in the bracket corresponds to fermion, gauge, Yukawa interaction and Higgs terms, respectively. $\phi$ is the physical higgs boson $m_{\phi}\simeq 125\GeV$. Similarly, we can get effective operators like
\begin{align*}
& m^2_\phi X^2 \phi^2,  X^2 F_{\mu\nu}F^{\mu \nu}, y X^2 \bar{\psi}\psi \phi, X^2 \bar{\psi}i\slashed{D}\psi, ... 
\end{align*}
with a common factor $\kappa^4 m_X^2/16\pi^2$. Again if $\mu_X \neq 0$ operators that induce DM decay would also arise,
\begin{align*}\label{eq:smdecay} 
X\bar{\psi}i\slashed{D}\psi, X F_{\mu\nu}F^{\mu \nu}, X D^\mu \phi D_\mu \phi,... ,
\end{align*}
with a common factor $\kappa^4\mu_X m_X^3/256\pi^4$. 

\section{Signatures of GDM}\label{sec:SGDM}

Now we discuss the possible signatures of GDM. If $Z^X_2$ symmetry is not broken, DM $X$ is stable but can pair-annihilate into other particles. The differential flux for particle $i$ is given by \begin{equation}\label{eq:flux}
	\frac{d\Phi_i}{dE}\sim\frac{1}{2}\frac{\langle \sigma v\rangle}{ m_X^2}\frac{d N_i }{dE}\int_0^{r_{c}} dr \rho ^2 \left(r\right),
\end{equation}
where $\langle \sigma v\rangle\sim \kappa^4m^2_X$ is averaged annihilation cross section, $d N_i/dE$ is energy spectrum for particle $i$, $r$ is the distance to galaxy center,  $r_c\sim 20$kpc for Milky Way and $\rho\sim \GeV/\textrm{cm}^3$ is DM density. We estimate the total flux is around $10^{-42}\textrm{cm}^{-2}\textrm{s}^{-1}$, which is independent of the DM mass. 
This flux seems too small to be probed by any known techniques. For example, current gamma-ray experiments are only sensitive to flux down to $10^{-4}\textrm{cm}^{-2}\textrm{s}^{-1}$ for TeV photons~\cite{Ibarra:2013cra}. It is therefore necessary to look for exotic astrophysical compact objects with high DM density $\rho\gtrsim 10^{20}\GeV/\textrm{cm}^3$, which might not be so surprising since we have already known the nucleon density can be as high as $10^{30}\GeV/\textrm{cm}^3$ in white dwarf stars and $10^{38}\GeV/\textrm{cm}^3$ in neutron stars, respectively.

When $Z^X_2$ symmetry is broken for $\mu_X\neq 0$, DM $X$ can decay. The lifetime of $X$ should be longer than the age of Universe, which puts a constraint on its dominant decay width $\Gamma^{h}_X$,
\begin{equation}\label{eq:upperbound}
\frac{m_X}{32\pi}\left[\frac{\mu_X m^2_X}{16\pi^2 M^2_P}\right]^2\lesssim t_U^{-1}
\textrm{ or } m_X\lesssim 10^{-11}\mu_X^{-\frac{2}{5}}M_P.
\end{equation}
If $\mu_X\simeq 1$, the upper bound for $m_X$ is $10^{7}\GeV$. The resulting flux is estimated as~\cite{Ko:2015nma}
\begin{equation}\label{eq:flux_g}
\frac{d\Phi_i}{dE}\sim \frac{\Gamma^i_X}{m_X}\frac{dN_i}{d E}\int_0^{r_{c}}dr \rho \left(r\right).
\end{equation}
Then the total flux of energetic graviton or high-frequency gravitational wave would be around $10^{-3}\textrm{cm}^{-2}\textrm{s}^{-1}\times\mu_X^2\times\left(m_X/10^7\GeV\right)^4$. So far no experiment searches for gravitons with such high energies. 

\begin{figure}[htb]
	\begin{center}
		\includegraphics[scale=0.6]{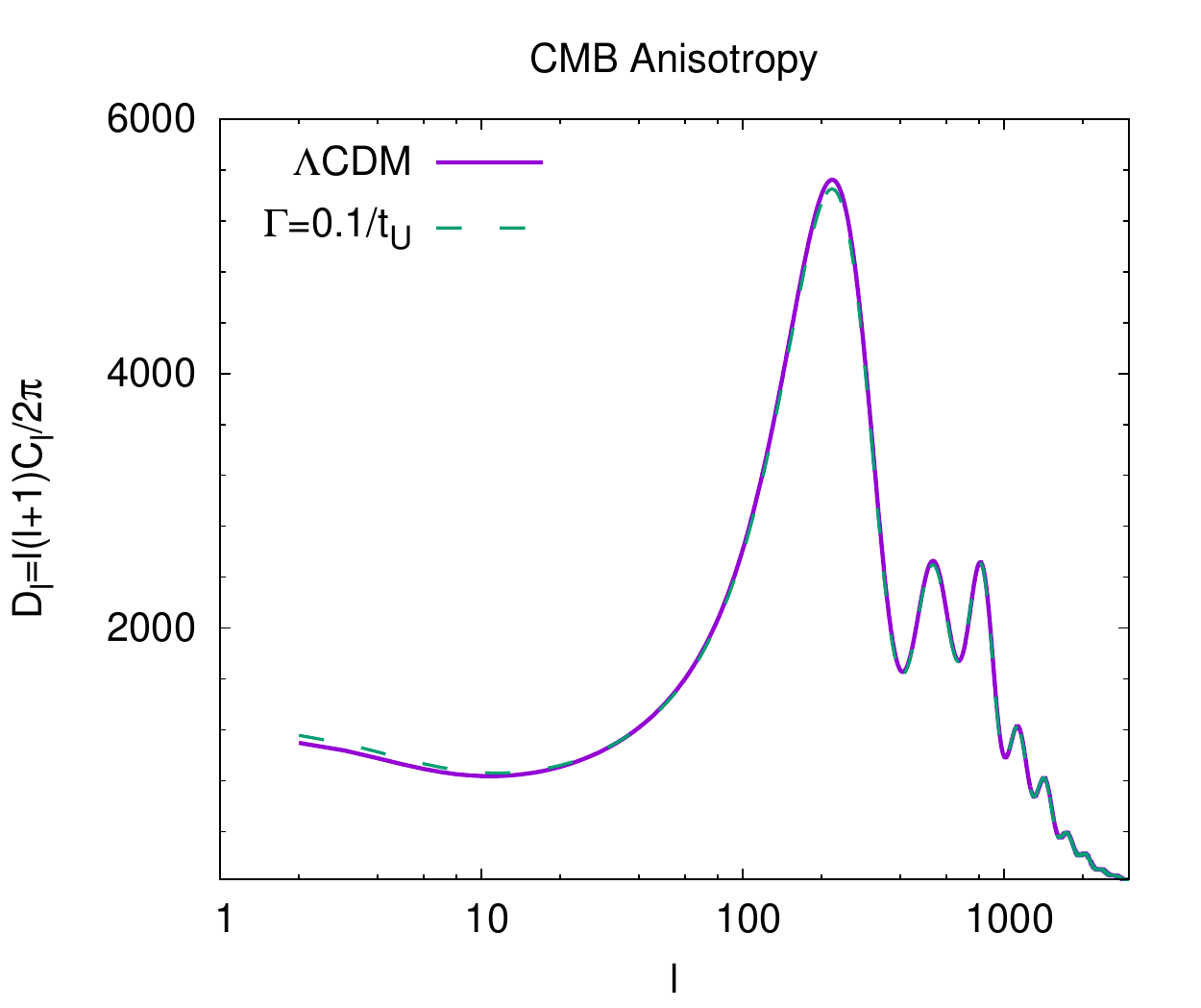}
		\caption{\label{fig:cmb}Effect on CMB temperature anisotropy from decaying DM $X$, illustrated with $\Gamma_X\sim 0.1t^{-1}_{U}$.}
	\end{center}
\end{figure}

However, decay of DM can also change the evolution of our late Universe by decreasing the matter component and increasing radiation part, which can be probed by CMB with enhanced late integrated Sachs-Wolfe effect at large scale or low $l$, as shown in Fig.~\ref{fig:cmb}. Current bound on decaying DM is $\Gamma_X\lesssim 0.1t^{-1}_{U}$~\cite{Audren:2014bca}. 

\begin{figure}[hbt]
	\begin{center}
		\includegraphics[scale=0.3]{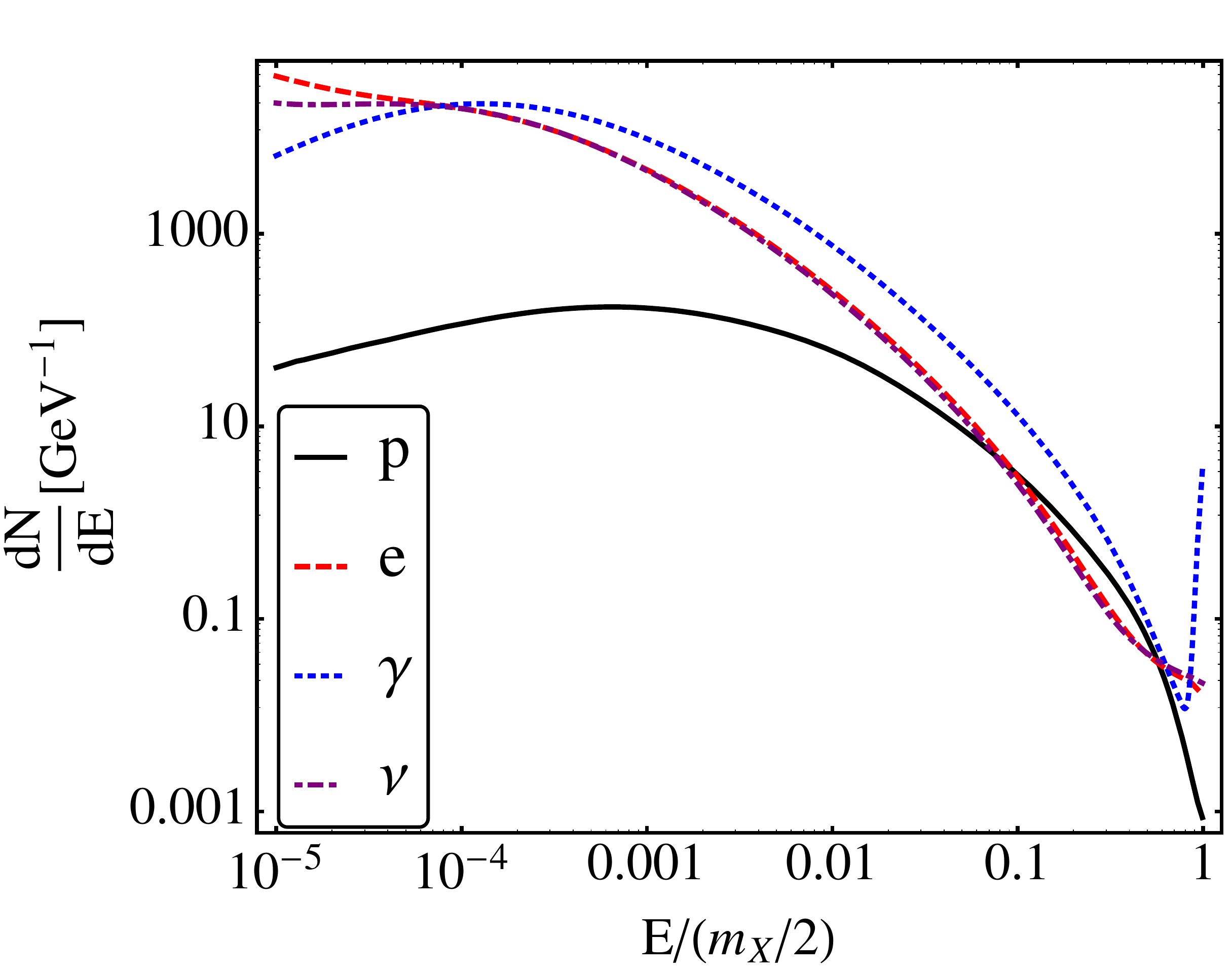}
		\caption{\label{fig:flux}Illustrations of spectra for $p,e^{\pm},\gamma,\nu$ from $X$'s decay with $m_X\gg\TeV$. A gamma-line component is presented at high end-point.}
	\end{center}
\end{figure}

GDM can also decay into to SM particles, such as  
%\begin{align}
$	X \rightarrow \phi  \phi, \gamma \gamma, ZZ, WW, gg, \psi  \bar{\psi}.$
%\end{align}
The partial decay width can be estimated as
\begin{equation}
\Gamma^{\textrm{SM}}_X \simeq \frac{m_X}{32\pi}\left[\frac{\mu_X m^4_X}{256\pi^4 M^4_P}\right]^2\sim \left[\frac{ m^2_X}{16\pi^2 M^2_P}\right]^2\Gamma^{h}_X .
\end{equation}
For $m_X \gg \TeV$, we also calculate the decay branch ratios 
\begin{align}
\mathcal{B}_{\phi\phi}:\mathcal{B}_{\gamma\gamma/ZZ}:\mathcal{B}_{WW}:\mathcal{B}_{gg}:\mathcal{B}_{\bar{\psi}\psi} \simeq 1:1:2:8:\frac{16m^2_\psi N_c}{m^2_X},\nonumber 
\end{align}
where $N_c=1,3$ for leptons and quarks, respectively. We then can make predictions for the spectra shapes of  $p,e^{\pm},\gamma,\nu$ from $X$ decay, as shown in Fig.~\ref{fig:flux}. Gamma line is also produced with energy $E=m_X/2$.
Unfortunately, the fluxes for SM particles are highly suppressed at least by a factor of $10^{-48}$ for $m_X\lesssim 10^7\GeV$, compared to graviton flux. These fluxes would be too small for searches in near future unless there are exotic astrophysical objects with very high density.

\section{Summary}\label{sec:summary}
In this paper, we have discussed a scenario that dark matter (DM) has only gravitational interactions. We have investigated how DM can be produced in the early universe and shown its mass range should be around $\TeV \lesssim m_X \lesssim 10^{11}\GeV$ to have the correct relic abundance. We have also considered DM decay if the discrete symmetry that protects DM's longevity is broken, and calculated the resulting graviton flux. Fluxes of other decaying products, like cosmic rays, gamma ray and neutrinos, are intrinsically very small, 
which is far below current experiments' sensitivity unless there are exotic astrophysical objects with very high DM density. Still, CMB can give indirect probe and constraint on the decay width of gravitational DM.

\acknowledgments
YT is partly supported by National Research Foundation of Korea Research Grant NRF-2015R1A2A1A05001869.

%\bibliographystyle{../JHEP}
%\bibliography{../references}

\begin{thebibliography}{1}
	
	\bibitem{Burgess:2003jk}
	C.~P. Burgess, 
	\emph{{Quantum gravity in everyday life}},
	\href{http://dx.doi.org/10.12942/lrr-2004-5}{\emph{Living Rev. Rel.} {\bf 7}
		(2004) 5--56}, [\href{http://arxiv.org/abs/gr-qc/0311082}{{\tt
			gr-qc/0311082}}].
	
	\bibitem{Donoghue:1994dn}
	J.Donoghue, 
	\emph{{GR as an effective field theory}},
	\href{http://dx.doi.org/10.1103/PhysRevD.50.3874}{\emph{Phys. Rev. }{\bf D50}(1994) 3874--3888}, [\href{http://arxiv.org/abs/gr-qc/9405057}{{\tt gr-qc/9405057}}].
			
	\bibitem{wu:2015wwa}
	Y.~L. Wu, \emph{{QFT of gravity with spin and scaling gauge invariance and spacetime dynamics with quantum inflation}}, Phys. Rev. {\bf D93} (2016) 024012, [\href{http://arxiv.org/abs/1506.01807}{{\tt 1506.01807}}].
      	
	\bibitem{Garny:2015sjg} 
	M.~Garny, M.~Sandora and M.~S.~Sloth,
	%``Planckian Interacting Massive Particles as Dark Matter,''
	Phys.\ Rev.\ Lett.\  {\bf 116}, 101302 (2016)
	[\href{http://arxiv.org/abs/1511.03278}{{\tt 1511.03278}}].      	
      	
	\bibitem{Ren:2014mta}
	J.~Ren and H.-J. He, % \emph{{Probing Gravitational Dark Matter}},
	\href{http://dx.doi.org/10.1088/1475-7516/2015/03/052}{\emph{JCAP}{\bf 1503}(2015)052}, [\href{http://arxiv.org/abs/1410.6436}{{\tt 1410.6436}}].
	
	\bibitem{Cata:2016dsg}
	O.~Catà, A.~Ibarra and S.~Ingenhütt, 
%	\emph{{Dark matter decays from non-minimal coupling to gravity}},
	[\href{http://arxiv.org/abs/1603.03696}{{\tt 1603.03696}}].
	
	\bibitem{Chung:1998ua} 
	D.~J.~H.~Chung, E.~W.~Kolb and A.~Riotto,
	%``Nonthermal supermassive dark matter,''
	Phys.\ Rev.\ Lett.\  {\bf 81}, 4048 (1998)
	[\href{http://arxiv.org/abs/hep-ph/9805473}{{\tt hep-ph/9805473}}].
	%%CITATION = doi:10.1103/PhysRevLett.81.4048;%%
	
	\bibitem{Kuzmin:1998kk} 
	V.~Kuzmin and I.~Tkachev,
	%``Matter creation via vacuum fluctuations in the early universe and observed ultrahigh-energy cosmic ray events,''
	Phys.\ Rev.\ D {\bf 59}, 123006 (1999)
	[\href{http://arxiv.org/abs/hep-ph/9809547}{{\tt hep-ph/9809547}}].
	%%CITATION = doi:10.1103/PhysRevD.59.123006;%%
	
	\bibitem{Khlopov}
	Our framework is also different from the so-called mirror particle where there exists a mirror copy of SM, see for example, 
	S.I.Blinnikov and M.Yu.Khlopov
	On the possible astronomical effects of `mirror' particles. Astron. Zh.
	(1983), V. 50, PP. 632-639. [Sov. Astron. (1983) V. 27,	PP. 371-355].
		
	\bibitem{book}
	E. W. Kolb and M. S. Turner, {\it The Early Universe} (Addison-Wesley, Reading, MA, 1990).
	
	\bibitem{Array:2015xqh} 
	P.~A.~R.~Ade {\it et al.} [BICEP2 and Keck Array],
	%``Improved Constraints on Cosmology and Foregrounds from BICEP2 and Keck Array Cosmic Microwave Background Data with Inclusion of 95 GHz Band,''
	Phys.\ Rev.\ Lett.\  {\bf 116}, 031302 (2016)
	[\href{http://arxiv.org/abs/1510.09217}{{\tt 1510.09217}}].
	
	\bibitem{Nurmi:2015ema} 
	Isocurvature perturbation may give some interesting constraints on the scalar case,
	S.~Nurmi, T.~Tenkanen and K.~Tuominen,
	%``Inflationary Imprints on Dark Matter,''
	\href{http://dx.doi.org/10.1088/1475-7516/2015/11/001}{JCAP {\bf 1511}, 001 (2015)},
	[\href{http://arxiv.org/abs/1506.04048}{{\tt 1506.04048}}].
	
	\bibitem{Ibarra:2013cra}
	A.~Ibarra, D.~Tran and C.~Weniger, 
	%	\emph{{Indirect Searches for Decaying Dark Matter}}, 
	\href{http://dx.doi.org/10.1142/S0217751X13300408}{\emph{Int. J.
			Mod. Phys.} {\bf A28} (2013) 1330040},
	[\href{http://arxiv.org/abs/1307.6434}{{\tt 1307.6434}}].
	
	\bibitem{Ko:2015nma}
	P.~Ko and Y.~Tang, 
   %	\emph{{IceCube Events from Heavy DM decays through the Right-handed Neutrino Portal}},
	\href{http://dx.doi.org/10.1016/j.physletb.2015.10.021}{\emph{Phys. Lett.}
		{\bf B751} (2015) 81--88}, [\href{http://arxiv.org/abs/1508.02500}{{\tt
			1508.02500}}].

	\bibitem{Audren:2014bca} 
	B.~Audren, J.~Lesgourgues, G.~Mangano, P.~D.~Serpico and T.~Tram,
	%``Strongest model-independent bound on the lifetime of Dark Matter,''
	JCAP {\bf 1412}, 028 (2014)
	[\href{http://arxiv.org/abs/1407.2418}{{\tt 1407.2418}}].
	%%CITATION = doi:10.1088/1475-7516/2014/12/028;%%
	
\end{thebibliography}
\providecommand{\href}[2]{#2}\begingroup\raggedright\endgroup
\end{document}